\documentclass[onecolumn,english,aps,floatfix,superscriptaddress,amsfonts,amssymb,nofootinbib, pra]{revtex4}
\usepackage{color}
\usepackage{comment}
\usepackage{graphicx}
\usepackage{amsmath}
\usepackage{latexsym}
\usepackage{slashed}
\usepackage{epstopdf}
\usepackage{amsmath,amssymb}
\usepackage{amssymb,amsmath}
\usepackage{multirow}
\usepackage{mathtools}
\usepackage[hidelinks]{hyperref}
\usepackage[dvipsnames]{xcolor}
\newcommand{\beq}{\begin{equation}}
\newcommand{\eeq}{\end{equation}}

\usepackage{tikz}
\usetikzlibrary{external}
\usepackage{xcolor}
\usetikzlibrary{patterns}
\usetikzlibrary{matrix,positioning,fit}
\usetikzlibrary{calc}
\newdimen\R
\begin{document}

\begin{center}{\large \textbf{
Scaling of the Formation Probabilities and Universal Boundary Entropies in the Quantum XY Spin Chain
}}\end{center}

\begin{center}
F. Ares\textsuperscript{1}, 
M.~A.~Rajabpour\textsuperscript{2}, 
J. Viti\textsuperscript{1,3,4},
\end{center}

\begin{center}
{\bf 1} {\small International Institute of Physics, UFRN, Campos Universit\'ario, Lagoa Nova 59078-970 Natal, Brazil}
\\
{\bf 2} {\small Instituto de F\'isica, Universidade Federal Fluminense, Av. Gal. Milton Tavares de Souza s/n, Gragoat\'a, 24210-346, Niter\'oi, RJ, Brazil}
\\
{\bf 3} {\small Escola de Ci\^encia e Tecnologia, UFRN, Campos Universit\'ario, Lagoa Nova 59078-970 Natal, Brazil}
\\
{\bf 4} {\small INFN, Sezione di Firenze, Via G. Sansone 1, 50019 Sesto Fiorentino, Firenze, Italy}

* viti.jacopo@gmail.com
\end{center}

\begin{center}
\today
\end{center}


 
 \begin{center} 
\bf{Abstract} 
\end{center}

{ 
We calculate exactly the probability to find the ground state of the XY chain in a given spin configuration in the transverse $\sigma^z$-basis. By determining  finite-volume corrections to the probabilities for a wide variety of configurations, we obtain the universal Boundary Entropy at the critical point. The latter is a benchmark of the underlying Boundary Conformal Field Theory characterizing each quantum state.
To determine the scaling of the probabilities, we prove a theorem that  expresses, in a factorized form, the eigenvalues of a sub-matrix of a circulant matrix as functions of the eigenvalues of the original matrix. 
Finally, the  Boundary Entropies are computed by exploiting a generalization of the Euler-MacLaurin formula to non-differentiable functions. It is shown that, in some cases, the spin configuration can flow to a linear superposition of Cardy states.
Our methods and tools are rather generic and can be applied to all the  periodic quantum chains which map to free-fermionic Hamiltonians.
}
 
\vspace{10pt}
\noindent\rule{\textwidth}{1pt}
\tableofcontents 
\noindent\rule{\textwidth}{1pt}
\vspace{10pt}

\section{Introduction}
The ground state of a quantum spin chain is usually a complicated state which, when  written on a local basis,
expands over an exponential number of terms. Each term in the expansion corresponds to a spin configuration on the selected basis.  Simultaneous measurements of all the local spins project the ground state into a single spin configuration with a certain probability, which is the absolute value squared of the overlap between such a state and the chosen configuration. The same is true for a  spinless  fermionic system on a lattice, where the measurement leads to a configuration whose lattice sites may or not be occupied by a fermion.

It is also possible to perform projective measurements on a subsystem. The simplest instance is perhaps the probability to observe the totality of the spins on a finite interval of the chain pointing up or down, which is the so-called Emptiness Formation Probability.  Such a quantity 
has been calculated analytically in a few integrable quantum chains, see \cite{Korepin:1994ui,Essler:1994se,Essler:1995vp,shiroishi2001emptiness,Kitanine2002a,
Korepin2003,FA,Painleve}. In the scaling limit, next to a critical point, the Emptiness Formation Probability  can be interpreted as statistical mechanics partition function on a cylinder or a strip with suitable boundary conditions~\cite{Stephan2013}. Within this formulation, it can be  studied by applying  Quantum Field Theory (QFT) and Conformal Field Theory (CFT) techniques, see also \cite{Rajabpour2015,Rajabpour2016,Viti2016}.
In particular, at criticality one can extract universal data such as the central charge \cite{Stephan2013} of the underlying CFT and the anomalous dimensions of all the scaling fields~\cite{Rajabpour2015}. A string of fully polarized spins  is however just one example among the  possible configurations that  can be fixed for the subsystem. The probability of finding a finite portion of the ground state in a generic spin configuration has been also
studied numerically in \cite{najafi2016formation,Najafi:2019ypm} and dubbed Formation Probability (FP). Similar connections to CFT can be  drawn for a wide variety of FPs~\cite{najafi2016formation}. 

FPs can be of course defined also for the whole spin chain, in this case they coincide with the absolute value squared of the ground state overlaps. 

Analogously, it is expected that at criticality the $O(1)$ correction to their large volume expansion is universal  and given by the Boundary Entropy (BE) introduced in~\cite{boundary}. For similar studies 
in the scaling limit  with integrability techniques, we refer to  \cite{LeClair1994,Dorey1998,Dorey2000,Friedan2004,Dorey2004,Pozsgay2010,Caetano2020}.
By determining the BEs,  one can infer the fixed point of the renormalization group flow, i.e. the conformal boundary state, attracting at large scales  any spin configurations~\cite{Cardy_Sci}.
Renormalization of the ground state of a perturbed CFT toward a conformal boundary state is also a  key assumption for the approach to non-equilibrium phenomena initiated in~\cite{CC}.  
Finite size corrections to the FPs for completely polarized states in the XY and XXZ chain have been discussed in
\cite{Wei2005,Stephan2009,Huang2010,Stephan2010} and \cite{Shi2010,Stephan2010b, SB} respectively. Overlaps  in gapless spin chain with central charge one, have been studied in~\cite{CS}. These analyses were  also relevant to understand whether geometric entanglement could serve as a  possible measure of multipartite entanglement \cite{Barnum2001,Wei2003}. 

Although the ground state overlaps---alias the FPs---seem  fundamental building blocks of a  quantum many-body theory, they have not been  extensively investigated.

In this paper, we try to make  the first steps toward a systematic study.  We  focus on the the quantum XY chain  and determine the  FPs in the transverse $\sigma^z$-basis for a wide variety of spin configurations.
After computing exactly their asymptotic behaviour for large volume, we are able to extract the subleading volume-independent contribution and point out the  boundary CFT characterizing each configuration.

Our methods and tools are quite generic and can be applied to any system that maps to  a periodic quadratic fermionic Hamiltonian. On the technical side, first,  we prove a novel result for circulant matrices which allows us to obtain a closed finite-size expression for the FPs.
Then we exploit the Euler-MacLaurin (EM) summation formulas to determine  their large volume expansion. To calculate the BE, in particular,
we recall a nice generalization of the EM famous theorem, which can be applied also to non-differentiable functions~\cite{Navot1,Navot2}.

The rest of the paper is organized as follows: in Sec.~\ref{sec2} we derive field theoretical predictions for the FPs in the XY chain and write down an explicit determinant representation for them; in Sec.~\ref{sec3}, we illustrate applications of the formalism to the fully polarized states and the N\'eel state; in Sec.~\ref{sec4}, FPs  are determined, together with their asymptotic behaviour in the large volume limit, for a wide class of states in the XY chain; in Sec.~\ref{sec5}, we focus on the XX chain and conclude in Sec.~\ref{sec6}. The paper has also three Appendices. Appendix~\ref{app} adapts the results of~\cite{Navot2} to the XY chain; Appendix~\ref{app2} contains a proof of the main technical novelty of this paper, namely a closed expression for the determinant of a sub-matrix of a circulant matrix.  Finally, Appendix~\ref{app3} adds some details to the examples examined in Sec.~\ref{sec4}.

\section{Formation Probabilities and Boundary Entropies in the XY chain}
\label{sec2}
\textit{Boundary Entropies at a Quantum Critical Point.---}We start by introducing the XY spin chain in a 
transverse field $h$, defined by the Hamiltonian~\cite{Lieb}
\begin{equation}
\label{Ham_XY}
H_{\text{XY}}=-\frac{1}{2}\sum_{n=1}^{L}\left[\frac{1+\gamma}{2}~\sigma^{x}_{n}\sigma^{x}_{n+1}+\frac{1-\gamma}{2}~\sigma^{y}_{n}\sigma^{y}_{n+1}
+h\sigma_{n}^{z}\right],
\end{equation}
where $\sigma_{n}^{\alpha}$ ($\alpha=1,2,3$) are Pauli matrices satisfying $[\sigma^{\alpha}_n,\sigma^{\beta}_m]=
2i\varepsilon^{\alpha\beta\nu}\delta_{n,m}\sigma^{\nu}_{n}$ and the parameter $\gamma$  is dubbed anisotropy. We furthermore assume periodic boundary conditions for the spins, namely
$\sigma_{n}^{\alpha}=\sigma_{n+L}^{\alpha}$, and restrict ourselves to  $L=2N$ even. 
The Hamiltonian in Eq.~\eqref{Ham_XY} commutes with the parity operator
\begin{equation}
\label{parity}
P=\prod_{n=1}^{L}\sigma_n^z,
\end{equation}
whose eigenvalues are $\mathcal{N}=\pm 1$. $P$ implements the $\mathbb Z_2$ symmetry of the model under spin flip in the $x$-direction.  The 
Hilbert space splits into a direct sum of two subspaces: that containing linear combinations of states with an odd number of 
down spins along the $z$-direction, the so-called Ramond (R) sector with $\mathcal{N}=-1$, and the one containing states with an 
even number of down spins, or Neveu-Schwarz (NS) sector, where $\mathcal{N}=1$.
Both subspaces have dimension $2^{L-1}$. In the region of the phase space $h^2+\gamma^2>1$, the ground state of the XY chain, which
will be denoted by $|\Omega\rangle$, belongs to the NS sector~\cite{Katsura, DR, OYNR} and we will examine only  this possibility
from now on. The ground state energy is
\begin{equation}
\label{gs_energy}
E_{\text{gs}}=-\frac{1}{2}\sum_{k=1}^{L}\varepsilon(\phi_k),
\end{equation}
where $\varepsilon(\phi)=\sqrt{(h-\cos \phi)^2+\gamma^2\sin^2\phi}$ and $\phi_k=\frac{2\pi}{L}(k-1/2)$ with $k=1,\dots, L$. Inside the 
circle $h^2+\gamma^2<1$, the lowest energy state oscillates between the R and NS sector and the analysis is more involved. In particular, 
the case $\gamma=0$ and $|h|<1$ will be discussed separately in Sec.~\ref{sec5}.   The energy gap of the XY chain~\cite{OYNR} closes 
as $O(L^{-1})$ along the critical lines $|h|=1$ and $\gamma\not=0$.  The low-energy  quasi-particle excitations are free Majorana 
fermions described by a CFT with central charge $c=1/2$: this is the Ising CFT. In the Majorana 
fermion language, the NS sector is spanned by states that contain only an even number of fermionic quasi-particles.

Take an element of the   $\sigma^z$-basis, $|\boldsymbol{\sigma}\rangle\equiv|\bullet_1\bullet_2\dots\bullet_{2N}\rangle$  
with $\bullet_n\in\{|\uparrow\rangle_n,|\downarrow\rangle_n\}$, an eigenvector of 
the local operators $\sigma_n^z$ associated to the eigenvalues $\pm 1$.  Consider then in the XY chain the amplitude
  \begin{equation}
  \label{quantum_pf}
 f_{\boldsymbol{\sigma}}(\beta,L)\equiv\langle\boldsymbol{\sigma}|e^{-\beta H_{\text{XY}}}|\boldsymbol{\sigma}\rangle,
 \end{equation} 
 in the limit $\beta\gg L\gg 1$; Eq.~\eqref{quantum_pf} could be also interpreted as the Return Amplitude~\cite{RA} analytically continued to imaginary times.
 \begin{figure}[t]
 \centering
 \includegraphics[width=0.334\textwidth]{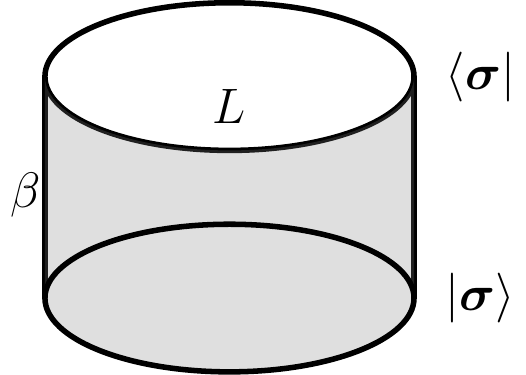}
\caption{The conformal partition functions on a annulus of circumference $L$ and height $\beta$. The state $|\boldsymbol{\sigma}\rangle$ acts as a boundary condition for the vertical imaginary time evolution. Rotating the annulus by $90$ degrees such a partition function can be interpreted as the conformal partition function of a system of finite length $\beta$ with temperature $1/L$. Imaginary time evolution happens now around the cylinder. In such a case the spins at edges should be described by a conformal invariant boundary condition~\cite{Cardyboundary}.}
\label{fig_an}
\end{figure}
Inserting a complete set of states, up to exponentially small corrections in the inverse temperature, one has
\begin{equation}
\label{limit}
f_{\boldsymbol{\sigma}}(\beta,L)\stackrel{\beta\gg L\gg 1}{\longrightarrow}|\langle\boldsymbol{\sigma}|\Omega\rangle|^2e^{-\beta E_{\text{gs}}},
\end{equation}
where $E_{\text{gs}}$ is the ground state energy in Eq.~\eqref{gs_energy}. For $L\gg 1$, the ground states overlap is expected to decay exponentially with a possible $O(1)$ term
\begin{equation}
\label{overlap}
\log|\langle\boldsymbol{\sigma}|\Omega\rangle|^2=-\Gamma_{\boldsymbol{\sigma}} L+2s_{\boldsymbol{\sigma}}+O(1/L),
\end{equation}
while extensivity of the ground state energy requires
\begin{equation}
\label{gs_exp}
E_{\text{gs}}=u L+u'-\frac{b}{L}+O(1/L^2).
\end{equation}
The coefficients $s_{\boldsymbol{\sigma}}$ and $b$ in the large volume expansions in Eqs.~(\ref{overlap}, \ref{gs_exp})  are dimensionless and argued to be universal, namely lattice-spacing independent, in the scaling limit. They can be calculated within a QFT formalism. Indeed,  the amplitude in Eq.~\eqref{quantum_pf} can be interpreted as a partition function on the annulus depicted in Fig.~\ref{fig_an}. At criticality, the bulk theory is conformal invariant and the state $|\boldsymbol{\sigma}\rangle$ acts as a boundary condition for the vertical imaginary time evolution. The latter is driven by the bulk conformal Hamiltonian with central charge $c$.
If the state $|\boldsymbol{\sigma}\rangle$ coincides with the ground state of a massive deformation of a CFT~\cite{Cardy_Sci}, at the bulk critical point, it renormalizes toward a conformal boundary state $|\Phi\rangle$.
In this case, the universal part of the quantum amplitude in Eq.~\eqref{quantum_pf}, in the limit $\beta\gg L\gg 1$, is given by~\cite{boundary}
\begin{equation}
\label{CFT_res}
f_{\boldsymbol{\sigma}}(\beta,L)\vert_{\text{univ}}\stackrel{\beta\gg L\gg 1}{\longrightarrow}g_{\Phi}^2~e^{\frac{\beta}{L}\frac{\pi c v_F}{6}},
\end{equation}
where $v_F$ is the Fermi velocity and $g_{\Phi}\equiv\langle\Phi|\Omega\rangle>0$ is the renormalized Boundary Entropy (BE)~\cite{boundary}.
In the following, we will examine  spin configurations $|\boldsymbol{\sigma}\rangle$ that exhibit a periodic pattern of period $p\ll L$ in real space and conjecture that for large volume they are still attracted by a conformal boundary state $|\Phi\rangle$. Our analysis, in particular, does not cover the possibility  of states that break translation invariance in the continuum limit, such as the domain wall $|\boldsymbol{\sigma}\rangle=|\overbrace{\uparrow\uparrow\dots\uparrow}^{L/2}\overbrace{\downarrow\dots\downarrow\downarrow}^{L/2}\rangle$ and which deserve a separate study~\cite{Ref1}. Eq.~\eqref{CFT_res} implies the celebrated~\cite{Cardyfree, Affleckfree} universality of the $O(1/L)$ term in the expansion of the ground state energy $b=(\pi c v_F)/6$, see Eq.~\eqref{gs_exp}.
For the critical XY chain, as long as $\gamma\not=0$, one has $v_F=\gamma$ and $c=1/2$; indeed the CFT result for $b$  can be readily checked applying the EM summation formula \eqref{EM1} to Eq.~\eqref{gs_energy}.

The notion of conformal boundary states, or Cardy states,  was introduced  in the seminal work~\cite{Cardyboundary} and nowadays is well established~\cite{Cardy_Rev}.
Here we only mention that in the Ising CFT there are two types of boundary  states, free  and fixed, distinguished by $\mathbb{Z}_2$ symmetry.
The free boundary state $|\Phi\rangle=|\text{free}\rangle$ has the property that $\langle\text{free}|\sigma_{n}^x|\text{free}\rangle=0$, while for fixed boundary states $|\Phi\rangle=|\pm\rangle$ one has $\langle\pm|\sigma_{n}^x|\pm\rangle=\pm 1$. However, if $\gamma\not=0$ and in absence of  a longitudinal field coupling to $\sigma_n^x$, all the states in the NS sector, when expressed in the $\sigma^x$-basis, are symmetric under $\sigma^x_{n}\rightarrow -\sigma_{n}^x$. Consequently, any spin configuration $|\boldsymbol{\sigma}\rangle$ in the NS sector should renormalize either to the free boundary state or to the linear superposition with equal weights of fixed boundary states.
In conclusion, field theory predicts that for $|h|=1$ the $O(1)$ term in Eq.~\eqref{overlap} does not depend even on $\gamma$ and is given by~\cite{Kon}
\begin{equation}
\label{predictions}
s_{\boldsymbol{\sigma}}(h)|_{h=\pm 1}=\log g_{\Phi}=\left\{\begin{array}{l}  0~\text{if $|\boldsymbol{\sigma}\rangle\stackrel{\text{flows to}}{\rightarrow}|\Phi\rangle=|\text{free}\rangle$}, \\
  \frac{1}{2}\log 2~\text{if $|\boldsymbol{\sigma}\rangle\stackrel{\text{flows to}}{\rightarrow}|\Phi\rangle=|+\rangle+|-\rangle $}.
\end{array}\right.
\end{equation}
The equal weight linear combination of fixed boundary states  in Eq.~\eqref{predictions} already occurred in the literature. For instance,  in the study of the boundary phase diagram of the Tricritical Ising model~\cite{Chim,afflecktri, GRW} and in the analysis of the renormalization group flow of the massive ground state of the Ising spin chain~\cite{Kon}.  As we will discuss at the end of Sec.~\ref{sec4}, the two boundary states $|+\rangle+|-\rangle$ and $|\text{free}\rangle$ are related by Kramers-Wannier (KW) duality~\cite{Kon}. The interpretation of the linear superposition in terms of a topological defect~\cite{GW, Kon2} and its appearance along a  boundary flow has also been recently emphasized in~\cite{FI}. Finally notice that in principle other $\mathbb Z_2$ symmetric linear superpositions of the fixed boundary states could appear in Eq.~\eqref{predictions}, leading to  larger boundary entropies. Nevertheless, as we will discuss in detail in the next sections, our results are consistent with a renormalization toward the simplest possibility given by the state $|+\rangle+|-\rangle$.
\newline

\textit{Determinant Representation for the Overlaps.---}
 We provide here  an explicit determinant representation for the overlap $\langle\boldsymbol{\sigma}|\Omega\rangle$ in the XY chain, see Eq.~\eqref{overlapXY}. This is the starting point of our study of the  FPs. Consider a state $|\boldsymbol{\sigma}\rangle$  with $2r$ down spins at positions: $1\leq i_1<i_2<\dots<i_{2r}\leq 2N$. By adapting to imaginary time the formalism in~\cite{RA},  the partition function  $f_{\boldsymbol{\sigma}}(\beta,L=2N)$ in Eq.~\eqref{quantum_pf} can be calculated as
\begin{equation}
\label{ret_amp}
 f_{\boldsymbol{\sigma}}(\beta,2N)=\frac{\text{Pf}(\textbf{M}_{\sigma}(\beta))}{\sqrt{\det(\textbf{Q}(\beta))}},
 \end{equation}
where the symbol $\text{Pf}$ denotes the Pfaffian ($[\text{Pf}(\textbf{A})]^2=\det\textbf{A}$ for $\textbf{A}$ antisymmetric). The antisymmetric matrix $\textbf{M}_{\sigma}$ is obtained from the $4N\times 4N$ antisymmetric matrix
\begin{equation}
\label{matrixM}
\textbf{M}=\begin{bmatrix}-i\textbf{X} & \textbf{Q}\\
            -\textbf{Q} & i\textbf{X}
           \end{bmatrix},\quad\textbf{Q}=\textbf{Q}^{\dagger},~\textbf{X}=\textbf{X}^{\dagger}
\end{equation}
by removing the columns  and rows $\{i_1,\dots, i_{2r}\}$ and  $\{i_1+2N,\dots, i_{2r}+2N\}$. The Hermitian matrices $\textbf{X}$ and $\textbf{Q}$ are circulant and commute; they are explicitly given by~\cite{RA}
\begin{align}
 &[\textbf{X}(\beta)]_{lm}=\frac{1}{2N}\sum_{k=1}^{2N}\frac{\gamma\sin(\phi_k)e^{-i\phi_k(l-m)}}{-h+\cos(\phi_k)+\varepsilon(\phi_k)\coth(\beta\varepsilon(\phi_k))}\\
 \label{matQ}
 &[\textbf{Q}(\beta)]_{lm}=\frac{1}{2N}\sum_{k=1}^{2N}\frac{e^{-i\phi_k(l-m)}}{\cosh(\beta\varepsilon(\phi_k))+\frac{-h+\cos(\phi_k)}{\varepsilon(\phi_k)}\sinh(\beta\varepsilon(\phi_k))}.
\end{align}
In order to determine the BEs, we shall evaluate them in the limit $\beta\rightarrow\infty$. From Eq.~\eqref{matQ} one obtains
\begin{equation}
\label{w_def}
\frac{1}{\sqrt{\det(\textbf{Q}(\beta))}}\stackrel{\beta\gg 1}{\longrightarrow}e^{\frac{1}{2}\beta\sum_{k=1}^{2N}\varepsilon(\phi_k)}\prod_{k=1}^{N}\left(\frac{1}{2}-\frac{h-\cos(\phi_k)}{2\varepsilon(\phi_k)}\right);
\end{equation}
the exponential prefactor above, cf. Eqs.~\eqref{gs_energy} and \eqref{limit}, reproduces $e^{-\beta E_{gs}}$ when $\gamma\not=0$. In the 
limit $\beta\rightarrow\infty$, 
on the other hand, $\textbf{Q}$ vanishes exponentially fast and   $\textbf{M}$  becomes block diagonal; we  can then define 
\begin{align}
\label{wlim}
 &[\textbf{W}]_{lm}\equiv\lim_{\beta\rightarrow\infty}[\textbf{X}(\beta)]_{lm}=\frac{1}{2N}\sum_{k=1}^{2N} e^{-i\phi_k(l-m)} w(\phi_k),\quad\text{with}\\
\label{w_eigen}
 & w(\phi_k)\equiv\frac{\gamma\sin(\phi_k)}{-h+\cos(\phi_k)+\varepsilon(\phi_k)}.
 \end{align}
From Eqs.~(\ref{ret_amp}-\ref{matrixM}) and Eq.~\eqref{limit}, it finally follows 
\begin{equation}
\label{overlapXY}
 |\langle\boldsymbol{\sigma}|\Omega\rangle|^2=\prod_{k=1}^{N}\left(\frac{1}{2}-\frac{h-\cos(\phi_k)}{2\varepsilon(\phi_k)}\right)|\det\textbf{W}_{\sigma}|,
\end{equation}
where $\textbf{W}_{\sigma}$ is the matrix extracted from $\textbf{W}$,  
by removing the columns and rows with indices $\{i_1,\dots, i_{2r}\}$ which are in correspondence with 
the positions of the down spins in the state $|\boldsymbol{\sigma}\rangle$. As a side 
remark, we observe that $\textbf{M}$ in Eq.~\eqref{matrixM} has the same formal structure as the correlation matrix 
derived in~\cite{FA} to evaluate the Emptiness Formation Probability, see also~\cite{Painleve}. After a few manipulations, 
its Pfaffian can be  rewritten as $\text{Pf}(\textbf{M})=|\det(\textbf{Q}+i\textbf{X})|$, from which Eq.~\eqref{overlapXY} also follows in the $\beta\rightarrow\infty$ limit. For similar determinant expressions, we refer to~\cite{Stephan2010}.

In Sec.~\ref{sec4}, see Appendix~\ref{app2} for a proof,  will be presented a formula for $\det\textbf{W}_{\sigma}$ valid for a large class of states,  which is particularly useful in  the large $N$ limit.

\section{Fully Polarized  States and the N\'eel State}
\label{sec3}
To validate the CFT predictions in Eq.~\eqref{predictions}, we start by analyzing FP for fully polarized states and the 
N\'eel state. Although the results presented in this Section are particular cases of the general discussion 
of Sec.~\ref{sec4}, we prefer to illustrate the main ideas first through these simpler examples. Results in 
this Section are valid for any $\gamma\not=0.$\\

\textit{Fully Polarized States.---} For the fully polarized up 
state, $|\boldsymbol{\sigma}\rangle=|\uparrow\dots\uparrow\rangle$, 
the matrix $\textbf{W}_{\uparrow\dots\uparrow}=\textbf{W}$. By applying directly Eq.~\eqref{overlapXY} we can then calculate the ground state overlap
\begin{equation}
\label{ov1}
\log|\langle\uparrow\dots\uparrow|\Omega\rangle|^2=\sum_{k=1}^{N}\log[ g_+(\phi_k)],
\end{equation} 
 with $g_{+}(\phi)=\frac{1}{2}+\frac{h-\cos(\phi)}{2\varepsilon(\phi)}$.
To compute the scaling with the system size of the FP, one shall apply the EM summation formula and approximate for large $N$ the sum in Eq.~(\ref{ov1}). The leading $O(N)$ contribution, cf. Eq.~\eqref{overlap}, is straightforward and $\Gamma_{\uparrow\dots\uparrow}=-\int_{0}^{\pi}\frac{d\phi}{2\pi}\log [g_{+} (\phi)]>0$.
However, contrary to the ground state energy in Eq.~\eqref{gs_energy}, the calculation of the subleading $O(1)$ term and therefore of the BE is  non-trivial. Indeed,  for some values of the parameters $(h,\gamma)$ the summand as a function of $\phi\in[0,\pi]$ is not differentiable and develops logarithmic singularities. 

\begin{figure}[t]
\centering
\includegraphics[width=0.5\columnwidth]{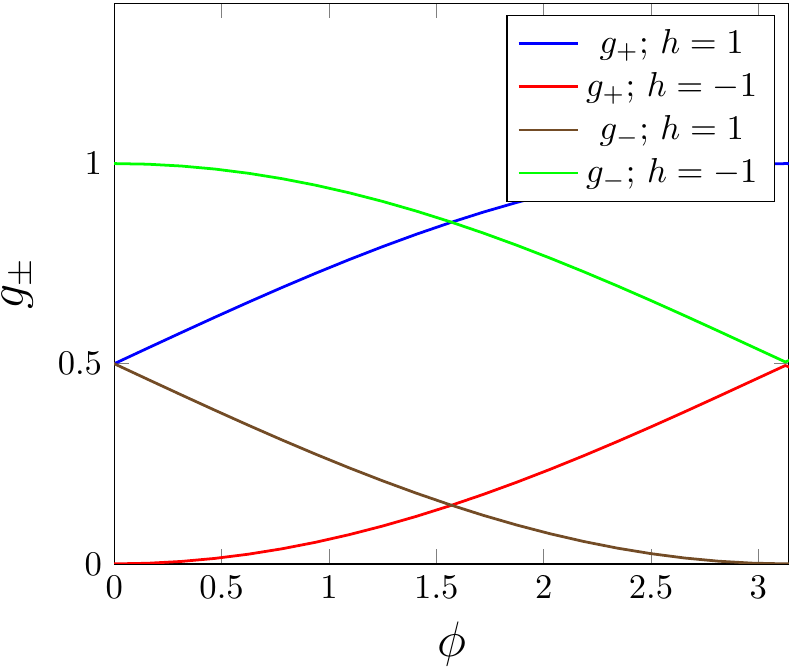}
\caption{The argument of the logarithms in Eqs.~(\ref{ov1}-\ref{ov2}) as a function of the angle $\phi\in[0,\pi]$ for $\gamma=1$ (Ising spin chain). $O(1)$ contributions in the large $N$ approximations of the sums in Eqs.~(\ref{ov1}-\ref{ov2}) are produced by zeros or divergences, according to Appendix~\ref{app}. For fully polarized states only zeros occur with $\alpha=2$.}
\label{fig_fully}
\end{figure}

Interestingly~\cite{Navot1, Navot2}, see Appendix~\ref{app}, logarithmic singularities are  responsible for the presence of non-zero $O(1)$ terms in the large $N$  expansion.
These, in turn, fix through Eq.~\eqref{overlap} the value of the BE along the critical lines $h=\pm 1$.
For instance, see also Fig.~\ref{fig_fully}, at $h=1$, the function $g_{+}$ is always positive and differentiable, while  vanishes quadratically at $\phi=0$,  when $h=-1$. In the former case $\log[g_+(\phi)]$ is a smooth function and the EM  summation formula~\eqref{EM1} gives $s_{\uparrow\dots\uparrow}(h)|_{h=1}=0$, corresponding to the free boundary state, cf. Eq.~\eqref{predictions}. In the latter, instead, $\log[g_+(\phi)]$ is singular at the boundary of the integration domain. The  extended EM summation formula~\eqref{EM2}  applies with $\alpha=2$ and leads to $s_{\uparrow\dots\uparrow}(h)|_{h=-1}=\frac{1}{2}\log 2$,  which indicates renormalization toward the linear superposition of fixed boundary states. Analogous considerations are valid for the fully polarized down state $|\boldsymbol{\sigma}\rangle=|\downarrow\dots\downarrow\rangle$. There is no matrix $\textbf{W}_{\downarrow\dots\downarrow}$ in Eq.~\eqref{overlapXY} and
\begin{equation}
\label{ov2}
 \log|\langle\downarrow\dots\downarrow|\Omega\rangle|^2=\sum_{k=1}^{N}\log [g_-(\phi_k)],
 \end{equation}
with $g_{-}|_{h}(\phi)=g_{+}|_{-h}(\pi-\phi)=\frac{1}{2}-\frac{h-\cos(\phi)}{2\varepsilon(\phi)}$, see also Fig.~\ref{fig_fully}. In particular, the values of the BE in a fully polarized down state as a function of the transverse field are 
reversed with respect to those of a fully polarized up state, that is $s_{\downarrow\dots\downarrow}(h)=s_{\uparrow\dots\uparrow}(-h)$. \\

\begin{figure}[t]
\centering
\includegraphics[width=0.5\columnwidth]{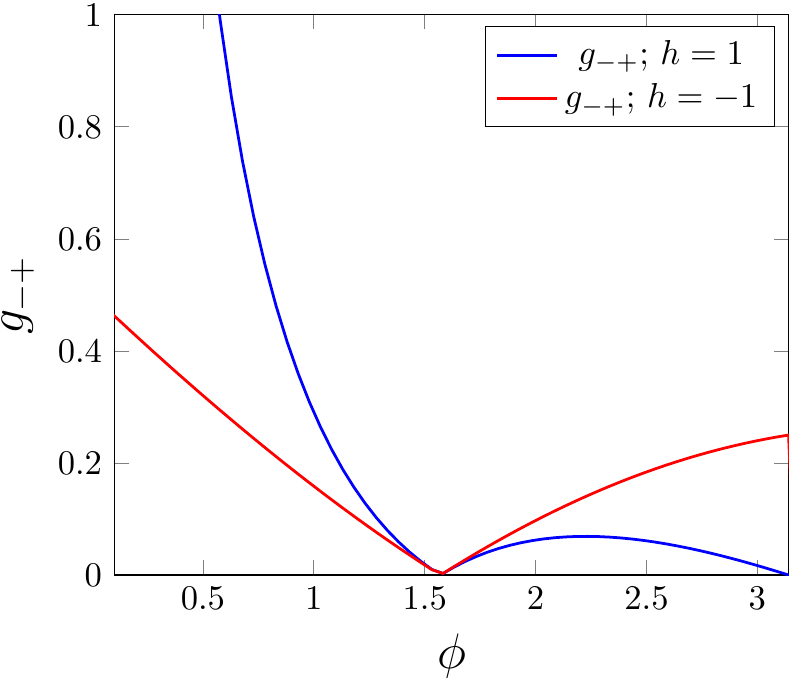}
\caption{The function $g_{-+}(\phi)$ plotted in the domain $\phi\in[0,\pi]$ at $\gamma=1$. In order 
to apply the extended EM summation formula in Eq.~\eqref{EM2}, we divide the interval $[0,\pi]$ in two, 
adding a boundary point at $\phi=\pi/2$. For $h=1$ (blue curve) one has a singularity with $\alpha=-1$ 
at $\phi=0$ and three additional singularities with $\alpha=1$, two at $\phi=\pi/2$ and one 
at $\phi=\pi$. When $h=-1$ instead (red curve), the function $g_{-+}$  has only a non-differentiable point at $\phi=\pi/2$. In both cases, the BE is $s_{\downarrow\uparrow\dots\downarrow\uparrow}=\frac{1}{2}\log2$, indicating a flow toward the linear combination of fixed boundary states.}
\label{fig_neel}
\end{figure}

\textit{N\'eel State.---} It is instructive to study separately also the N\'eel state, $|\boldsymbol{\sigma}\rangle=|\downarrow\uparrow\dots\downarrow\uparrow\rangle$; the state will be in the NS sector if $N$ is even; i.e. if the total length $L=2N$ of the chain is divisible by four. The matrix $\textbf{W}_{\sigma}$, in Eq.~\eqref{overlapXY}, is obtained by removing the odd columns and rows from \textbf{W} in Eq.~\eqref{wlim}. Since the smaller matrix is circulant, its eigenvalues can be computed by elementary means and from Eq.~\eqref{overlapXY} one obtains
\begin{align}
\label{Neel1}
 &\log|\langle\downarrow\uparrow\dots\downarrow\uparrow|\Omega\rangle|^2=\sum_{k=1}^{N}\log[g_{-+}(\phi_k)],\quad\text{with}\\
 \label{Neel2}
 &g_{-+}(\phi)=\frac{\gamma\sin(\phi)}{4\varepsilon(\phi)}\left|1-\frac{\varepsilon(\phi)-h+\cos(\phi)}{\varepsilon(\phi+\pi)-h-\cos(\phi)}\right|.
\end{align}
 The leading term in the large $N$ expansion of Eq.~\eqref{Neel1} is $\Gamma_{\downarrow\uparrow\dots\downarrow\uparrow}=-\int_{0}^{\pi}\frac{d\phi}{2\pi}\log[g_{-+}(\phi)]$, which can be shown to be positive by direct numerical integration. The value of the $O(1)$ contribution is again set by the zeros and the singularities of the function $g_{-+}(\phi)$ with $\phi\in[0,\pi]$. 
 For $h=1$,  $g_{-+}(\phi)$ diverges as $1/\phi$ close to $\phi=0$, is not differentiable at $\phi=\pi/2$ and vanishes linearly at $\phi=\pi$. According to Appendix~\ref{app} and Eq.~\eqref{overlap}, one then obtains $s_{\downarrow\uparrow\dots\downarrow\uparrow}(h)|_{h=1}=\frac{1}{2}\log 2$, indicating renormalization toward the linear combination of fixed boundary states. For $h=-1$, $g_{-+}$ has  only a non-differentiable point at $\phi=\pi/2$, leading again to a BE $s_{\downarrow\uparrow\dots\downarrow\uparrow}(h)|_{h=-1}=\frac{1}{2}\log 2$, consistent with the symmetry under spin flip in the $z$-direction of the N\'eel state. The function $g_{+-}(\phi)$ is plotted for $h=\pm 1$ and $\gamma=1$ in Fig.~\ref{fig_neel}; the caption provides a few additional details on the application of Eq.~\eqref{EM2}.
\section{Formation Probabilities and Boundary Entropies for Generic Spin Configurations}
\label{sec4}
We now present a rather general technique to calculate  the FP of  eigenstates $|\boldsymbol{\sigma}\rangle$ of the local spin operators $\sigma_n^z$. This is based, see Eqs.~(\ref{c_coeff_full}-\ref{matA}), on a  factorized expression for their overlap with the XY ground state.

Along the lines of~\cite{LC}, we discuss states that are obtained by repeating an elementary block $\mathcal{B}_{s,p}$ of $p$ spins; 
inside any block there are $s$ consecutive up spins. 
Conventionally and without loosing in generality, all the 
states except the fully polarized up state (i.e. the block $\mathcal{B}_{1,1}$) start with a down spin. For instance the N\'eel state, discussed in the previous Section, is 
labelled by the block $\mathcal{B}_{1,2}$; analogously a state 
such as $|\downarrow\downarrow\uparrow\uparrow\dots\downarrow\downarrow\uparrow\uparrow\rangle$ is in 
correspondence with the block $\mathcal{B}_{2,4}$. Defining
\begin{equation}
\label{M-def}
M\equiv\frac{2N}{p},
\end{equation}
one then finds a total of $sM$ up spins at positions $q=jp+r$, with $j=0,\dots,M-1$ and $r=p-s+1,\dots,p$;  we can 
further ensure that these states belong to the NS sector choosing $N$ a  multiple of $p$. To determine the FPs one shall calculate the determinant in Eq.~\eqref{overlapXY}; in this regard,  in Appendix~\ref{app2}, we will prove the following formula  
\begin{equation}
\label{c_coeff_full}
 \det\textbf{W}_{\sigma}=\prod_{k=1}^M~\mathcal{P}_0\left(w(\phi_k),w(\phi_{k+M}),\dots,w(\phi_{k+(p-1)M})\right),
\end{equation}
with $w(\phi)$ given in Eq.~\eqref{w_eigen}. $\mathcal{P}_0(x_1,\dots,x_p)$ in Eq.~\eqref{c_coeff_full} is a polynomial  
which coincides with the first non-vanishing coefficient  ---that of the power of degree $(p-s)$--- of the 
characteristic polynomial of the $p\times p$ matrix
\begin{equation}
\label{matA}
 [\textbf{A}]_{lm}=\frac{x_l}{p}\left[s~\delta_{l,m}-(1-\delta_{l,m})e^{\frac{i\pi (l-m)(p-s)}{p}}\frac{\sin\left(\frac{\pi(l-m)(p-s)}{p}\right)}{\sin\left(\frac{\pi(l-m)}{p}\right)}\right];\quad l,m=1,\dots,p.
\end{equation}
Eqs.~\eqref{c_coeff_full} and~\eqref{overlapXY} can be then used to calculate analytically 
the ground state overlaps of the  states $|\boldsymbol{\sigma}\rangle$, labelled by the block $\mathcal{B}_{s,p}$, as
\begin{equation}
\label{overlap_fin}
\log\left|\frac{\langle\boldsymbol{\sigma}|\Omega\rangle}{\langle\downarrow\dots\downarrow|\Omega\rangle}\right|^2=\sum_{k=1}^{M}\log\left|\mathcal{P}_0(w(\phi_k), w(\phi_{k+M}),\dots,w(\phi_{k+M(p-1)}))\right|.
\end{equation}
The coefficient $\Gamma_{\boldsymbol{\sigma}}$ in Eq.~\eqref{overlap} ruling the leading large $N$ behaviour of the FPs is then
\begin{equation}
 \Gamma_{\boldsymbol{\sigma}}=\Gamma_{\downarrow\dots\downarrow}+\int_{0}^{\frac{2\pi}{p}}\frac{d\phi}{2\pi}\log\bigl|\mathcal{P}_0\bigl(w(\phi),w(\phi+2\pi/p),\dots,w(\phi+2\pi(p-1)/p)\bigr)\bigr|,
\end{equation}
and from Eq.~\eqref{overlap_fin}, the BEs are also determined in analogy to what was done in Sec.~\ref{sec3} for 
the fully polarized and the N\'eel states.
We could also extend the analysis of the $O(1)$ term in the large $N$ expansion of Eq.~\eqref{overlap_fin} for 
any points outside the circle $h^2+\gamma^2=1$.  Notice that, by symmetry,  if $|\boldsymbol{\sigma}'\rangle$ is 
obtained by flipping in the $z$-direction all the spins of $|\boldsymbol{\sigma}\rangle$, it must 
hold $s_{\boldsymbol{\sigma}}(h)=s_{\boldsymbol{\sigma}'}(-h)$. The property is shared, for example, by the state $|\boldsymbol{\sigma}\rangle$, 
labelled by the block $\mathcal{B}_{s,p}$,  and  its companion $|\boldsymbol{\sigma'}\rangle$, labelled 
by  $\mathcal{B}_{p-s,p}$. Its verification provides a non-trivial test of the formalism.

Based on a case by case study which is reported below and in Appendix~\ref{app3}, a pattern emerges for the BEs that will be illustrated at the end of this Section, together with a physical interpretation. \\

\textit{Example 1.---} Consider the state associated to the block $\mathcal{B}_{s=1, p}$; this is of 
the form $|\overbrace{\downarrow\dots\downarrow}^{p-1}\uparrow\dots\rangle$. 
For $s=1$, the polynomial $\mathcal{P}_{0}(x_1,\dots,x_p)$  is minus the trace of the matrix $\textbf{A}$ in Eq.~\eqref{matA}; namely 
\begin{equation}
\label{pol_p}
 \mathcal{P}_0(x_1,\dots,x_p)=-\frac{1}{p}\sum_{j=1}^{p}x_j.
\end{equation}
From Eq.~\eqref{overlap_fin}, one obtains the ground state overlap for the class of states labelled by the block $\mathcal{B}_{1,p}$
\begin{equation}
\label{overlap-p}
 \log\left|\frac{\langle\boldsymbol{\sigma}|\Omega\rangle}{\langle\downarrow\dots\downarrow|\Omega\rangle}\right|^2=\sum_{k=1}^{2N/p}\log[g_{1,p}(\phi_k)].
\end{equation}
 The function $g_{1,p}$, cf. Eq.~\eqref{pol_p}, reads
\begin{equation}
\label{gp}
 g_{1,p}(\phi)=\frac{1}{p}\left|\sum_{j=0}^{p-1}w\left(\phi+\frac{2\pi j}{p}\right)\right|,
\end{equation}
where we have further used  $\phi_{k+m}=\phi_{k}+\frac{\pi m}{N}$ for any integer $m$. The same result could be derived 
more prosaically observing that for $s=1$  the matrix $\textbf{W}_{\sigma}$ in Eq.~\eqref{finalc}  is circulant and  
its eigenvalues can be calculated directly.

\begin{figure}[t]
\centering
\includegraphics[width=0.5\columnwidth]{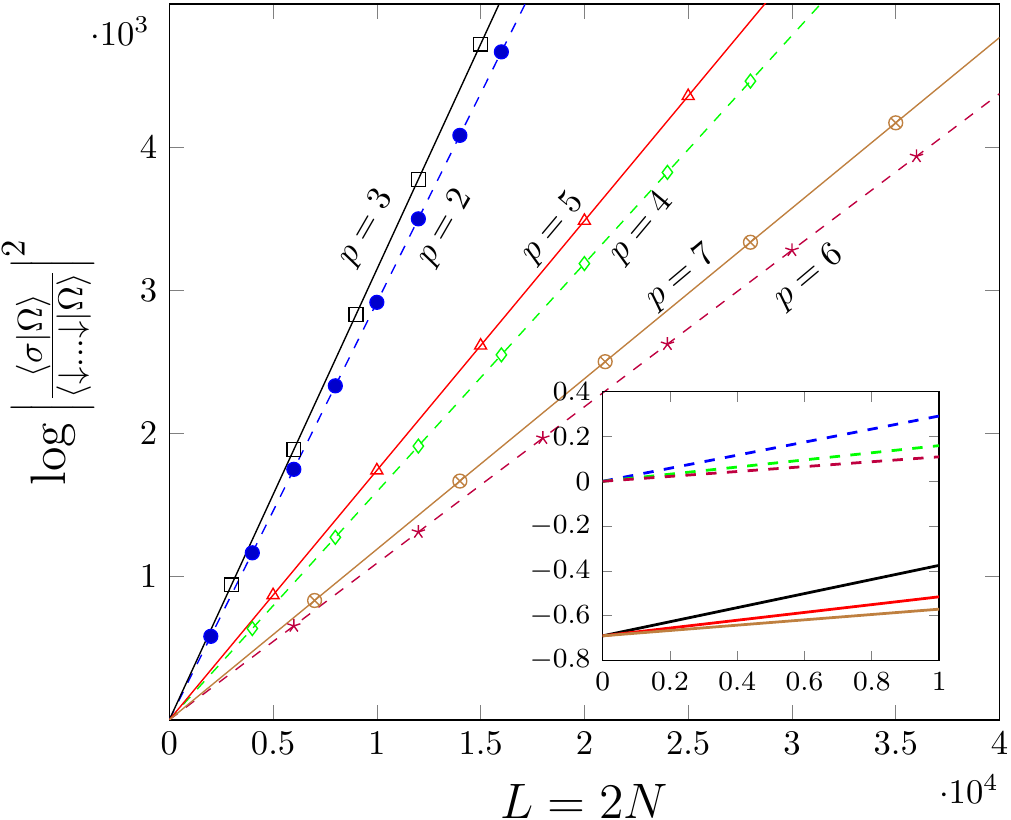}
\caption{Direct evaluation of the sums in Eq.~\eqref{overlap-p} for several values of $p$ for $h=1$ and $\gamma=1$. The straight 
lines are linear fits of the data obtained at different $L=2N$. The inset contains the behaviour of the straight 
lines close to the origin: the $O(1)$ term in large $N$ expansion of Eq.~\eqref{overlap-p} is $-\log 2$ for odd $p$ (solid lines) 
and zero for even $p$ (dashed lines). These values follow from application of Eq.~\eqref{EM2} and the properties of the 
function $g_{1,p}(\phi)$ in Eq.~\eqref{gp}.}
\label{fig_un}
\end{figure}

As explained in Sec.~\ref{sec3}, by identifying the zeros and the singularities in the interval $\phi\in[0,2\pi/p]$ of  $g_{1,p}$ in Eq.~\eqref{gp} 
we can determine the BEs. We start from the line $h=1$. For $p$ even, the function $g_{1,p}(\phi)$ diverges at $\phi=0$ and $\phi=2\pi/p$, while 
is vanishing and non-differentiable at $\phi=\pi/p$. According to Eq.~\eqref{EM2} there is no $O(1)$ term when approximating the 
sum in Eq.~\eqref{overlap-p} for large $N$. By combining this result with the analysis of the analogous 
contribution coming from $\log|\langle\{\downarrow\}|\Omega\rangle|^2$ in Eq.~\eqref{overlap-p}, see Sec.~\ref{sec3}, 
we conclude that  states associated to configurations $\mathcal{B}_{1,p} $ and $p$ even  renormalize toward the 
linear superposition of fixed boundary states. Interestingly when $p$ is odd,  this is no longer the case. For $p$ 
odd and $h=1$, the function $g_{1,p}(\phi)$ has only a pole at $\phi=\pi/p$. In the notations of Appendix~\ref{app}, 
there are then two singularities with $\alpha=-1$ and
the $O(1)$ term in the large $N$ expansion of Eq.~\eqref{overlap-p} has value  $-\log 2$. Taking into account the 
contribution of the fully polarized down state, it follows  that states with  $p$ odd renormalize toward the free boundary state. These 
findings are also summarized in Fig.~\ref{fig_un}, which illustrates  the universality of the BEs along the quantum critical line $h=1$.

Finally, at $h=-1$, the function $g_{1,p}(\phi)$ for both even and odd $p$ has two singularities with $\alpha=1$ in 
the domain $\phi\in[0,2\pi/p]$. In such a case, there is no $O(1)$ term 
in Eq.~\eqref{overlap-p} coming from the fully polarized down state.  We then conclude that the 
states associated to $\mathcal{B}_{1,p}$ have BE $s_{\sigma}=\frac{1}{2}\log 2$ at $h=-1$, indicating 
renormalization toward the linear superposition of fixed boundary states for any $p>1$.\newline

\textit{Example 2.---} Consider now the more general states associated to the blocks $\mathcal{B}_{s,p}$ for $s>1$. 
For the sake of brevity,  we sketch out two examples with $p=4$ and defer to Appendix~\ref{app3} the rest of the analysis.  We first focus on $s=2$ and $s=3$; in this case, the blocks $\mathcal{B}_{2,4}$ and  $\mathcal{B}_{3,4}$ label the states $|\downarrow\downarrow\uparrow\uparrow\dots\downarrow\downarrow\uparrow\uparrow\rangle$ and  $|\downarrow\uparrow\uparrow\uparrow\dots\downarrow\uparrow\uparrow\uparrow\rangle$ respectively. When $s=2$, the polynomial $\mathcal{P}_0(x_1,\dots,x_4)$ in Eq.~\eqref{c_coeff_full} is
\begin{equation}
 \mathcal{P}_0(x_1,x_2,x_3,x_4)=\frac{1}{8}\left[x_1(x_2+2x_3+x_4)+x_2(x_3+2x_4)+x_3x_4\right],
\end{equation}
and substituting into  Eq.~\eqref{overlapXY} with $M=N/2$ one obtains the ground state overlap. The analysis of the zeros and singularities of $g_{2,4}(\phi)\equiv|\mathcal{P}_0\bigl(w(\phi), w(\phi+\pi/2), w(\phi+\pi), w(\phi+3\pi/2)\bigr)|$ in the interval $\phi\in[0,\pi/2]$ reveals that the state  $|\downarrow\downarrow\uparrow\uparrow\dots\downarrow\downarrow\uparrow\uparrow\rangle$ renormalizes to the free boundary state for both $h=\pm 1$. Curiously  this is the opposite behaviour of the N\'eel state. For $s=3$ the polynomial entering Eq.~\eqref{overlap_fin} is instead
\begin{equation}
 \mathcal{P}_0(x_1,x_2,x_3,x_4)=-\frac{1}{4}(x_1x_2x_3+x_1x_2x_4+x_1x_3x_4+x_2x_3x_4),
\end{equation}
to which is associated the function $g_{3,4}(\phi)$ in complete analogy with the case $s=2$. By analyzing zeros and singularities of $g_{3,4}(\phi)$ for $\phi\in[0,\pi/2]$ it is possible to conclude that the state $|\downarrow\uparrow\uparrow\uparrow\dots\downarrow\uparrow\uparrow\uparrow\rangle$ renormalizes to the linear combination of fixed boundary states for both $h=\pm 1$.  It is also easy to verify that $s_{\downarrow\downarrow\downarrow\uparrow\dots\downarrow\downarrow\downarrow\uparrow}(h)=s_{\downarrow\uparrow\uparrow\uparrow\dots\downarrow\uparrow\uparrow\uparrow}(-h)$ for $h\in\mathbb R$, consistently with the discussion below Eq.~\eqref{overlap_fin}.\\

\textit{A general pattern and the KW duality.---}The analysis carried out in the examples above and in Appendix~\ref{app3} 
is consistent with the following pattern: When the block $\mathcal{B}_{s,p}$ contains an even number of down spins (i.e. $p-s$ is even) the 
state $|\boldsymbol{\sigma}\rangle$ flows to the free boundary state at $h=1$; otherwise to the linear superposition of 
fixed boundary states. Along the line $h=-1$, the same is true if the block $\mathcal{B}_{s,p}$ contains an even 
number of up spins (i.e. $s$ is even). At present, we do not have a  formal proof of this statement but we can provide a physical interpretation for the critical Ising spin chain, based on the KW duality. In short, by exploiting the KW duality, one can  infer the  renormalization flow of eigenstates of the $\sigma^z$-basis by mapping them into eigenstates of the dual spin basis, which is isomorphic with the $\sigma^{x}$-basis.

To be definite, let us  consider Eq.~\eqref{Ham_XY} for $\gamma=1$ and $h=1$. In the  KW mapping one introduces the dual spin variables $\mu^{\alpha}_l$ on  the edges of the chain through
\begin{equation}
\label{dual}
 \mu_{n+1/2}^z=\eta\prod_{j=1}^n\sigma_{j}^z,\quad~\mu_{n+1/2}^{x}=\sigma_{n}^x\sigma_{n+1}^x,
\end{equation}
for $n=0,\dots,L-1$ and $\eta=\pm 1$.  Notice that in the NS sector $\mu^{z}_{L+1/2}=\eta$; therefore, when working in the $\mu^z$-basis,  we shall  fix  $\eta=\mu_{1/2}^{z}$ to guarantee that also the dual Hamiltonian will have periodic boundary conditions in the dual spin variables.
Nevertheless, because the operator $\prod_{n=1}^{L}\mu_{n-1/2}^{x}$ acts as the identity on the dual Hilbert space, the latter  has still dimension $2^{L-1}$ and is spanned by all the $\mathbb Z_2$ even  states in the $\mu^{z}$-basis.  With this \textit{caveat}, the Ising chain Hamiltonian restricted to the NS sector after the KW mapping reads
\begin{equation}
 H=-\frac{1}{2}\sum_{n=1}^{L}\left[\mu_{n-1/2}^{x}+\mu^{z}_{n-1/2}\mu^{z}_{n+1/2}\right],~~\quad \mu_{n}^{\alpha}=\mu_{n+L}^{\alpha},
\end{equation}
which  is the same as the original one upon exchanging the  transverse and longitudinal degrees of freedom. A similar treatment of the KW duality is also contained in~\cite{R19}.

We can now investigate how the KW duality transforms the Hilbert spaces; eigenstates of $\mu^z$ with eigenvalues $\pm 1$ will be denoted by $|\rightarrow\rangle$ and $|\leftarrow\rangle$ respectively. By recalling the definition of the dual spin in Eq.~\eqref{dual}, one can conclude that 
\begin{equation}
\label{KW_sym}
 |\uparrow\dots\uparrow\rangle\stackrel{\text{maps into}}{\longleftrightarrow} |\rightarrow\dots\rightarrow\rangle+|\leftarrow\dots\leftarrow\rangle.
\end{equation}
If $h=1$, a state fully polarized in the positive $z$-direction corresponds to the free boundary state~\cite{Stephan2013}, and the KW transformation in Eq.~\eqref{KW_sym}  maps it  into the linear superposition of  fixed boundary states~\cite{Kon, FI, chiral}. Such a result was anticipated at the end of Sec.~\ref{sec2}. Analogously when applying the  mapping to the state $|\downarrow\dots\downarrow\rangle$ it turns out
\begin{equation}
\label{duality2}
 |\downarrow\dots\downarrow\rangle\stackrel{\text{maps into}}{\longleftrightarrow}|\rightarrow\leftarrow\rightarrow\leftarrow\dots\rangle+|\leftarrow\rightarrow\leftarrow\rightarrow\dots\rangle,
\end{equation}
which is the $\mathbb Z_2$ symmetric N\'eel state in the dual spin basis. Under coarse-graining of the dual spins, for instance by decimation, the N\'eel state on the RHS of the duality should flow to the free boundary state. This implies that the fully polarized down state on the LHS of the duality   flows instead to the linear superposition of fixed boundary states.  As a last example consider the state $|\uparrow\uparrow\downarrow\downarrow\dots\rangle$; the KW transformation acts as
\begin{equation}
\label{duality3}
|\uparrow\uparrow\downarrow\downarrow\dots\rangle\stackrel{\text{maps into}}{\longleftrightarrow}|\rightarrow\rightarrow\rightarrow\leftarrow\dots\rangle+|\leftarrow\leftarrow\leftarrow\rightarrow\dots\rangle.
\end{equation}
Under coarse-graining of the dual spins the RHS of Eq.~\eqref{duality3} renormalizes to linear superposition of fixed boundary states, therefore the LHS will flow to the free boundary state.

\section{Formation Probabilities and Boundary Entropies for the XX Chain}
\label{sec5}

\begin{figure}[t]
\centering
 \includegraphics[width=0.94\textwidth]{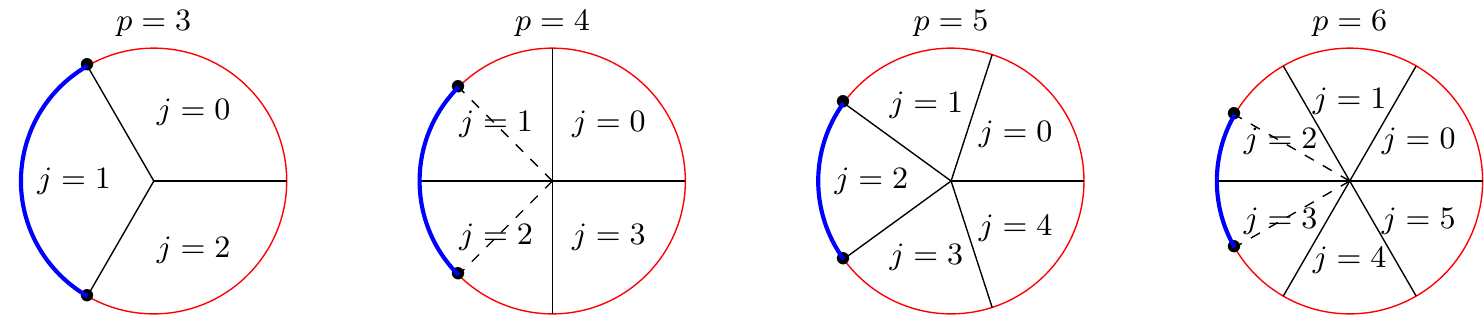}
 \caption{The figure shows how the indices $\boldsymbol{j}=\{j_1,\dots,j_{2N/p}\}$ must  be chosen in 
 order to minimize the $\lambda$'s in Eq.~\eqref{sum0}. For $\phi_k\in(0,2\pi/p)$, the angles $\phi_{k}+2\pi j_k/p$  
 fall into the sectors indicated in the figure depending on the value of $j_k$. It is clear that to 
 minimize $\lambda(\phi_k)$ we should choose $\bar{j}_k$ such that $\phi_{k}+2\pi \bar{j}_k/p$ belongs 
 to the blue arc $(\pi-\pi/p,\pi+\pi/p)$. In particular, if $p=2m+1$, $\bar{j}_k=m$ for any $\phi_k$, 
 while if $p=2m$, $\bar{j}_k=m-1$ when $\phi_k>\pi/p$ and $\bar{j}_k=m$ if $\phi_k<\pi/p$. There is only 
 one choice of $\boldsymbol{j}$ that realizes the minimum; the blue arc is a Fermi sea for $\arccos(h)=\pi-\pi/p$. }
 \label{fig0}
 \end{figure}

Along the line $\gamma=0$, the $\mathbb Z_2$ symmetry under spin reversal (in the $x$-direction) of 
the XY spin chain is promoted to a $U(1)$ symmetry conserving the total magnetization $\mathfrak{M}=\sum_{n=1}^{2N}\sigma_{n}^z$ in the $z$-direction.
The XY Hamiltonian in Eq.~\eqref{Ham_XY} can be easily diagonalized  as
\begin{equation}
\label{ham_gamma}
 H_{\text{XY}}|_{\gamma=0}=\sum_{k=1}^{2N}\lambda(\phi_k)d^{\dagger}_kd_k+\frac{hL}{2};
\end{equation}
the $d$-operators satisfy $\{d^{\dagger}_k,d_{k'}\}=\delta_{k,k'}$ and $\lambda(\phi)=-h+\cos(\phi)$.
In a magnetization sector $\langle\mathfrak{M}\rangle$, the ground state energy is obtained by filling all 
the single-particle negative energy levels, i.e. the low energy state is a Fermi sea. More precisely, 
if $\bar{k}\in\mathbb R$ is such that $\phi_{\bar{k}}=\arccos(h)$ and $\mathcal{S}(h)$ is the set of 
integers $\mathcal{S}(h)\equiv\{\lceil \bar{k}\rceil,\dots, 2N-\lfloor \bar{k}\rfloor\}$ then the ground state energy is
\begin{equation}
\label{gs_free}
 E_{\text{gs}}|_{\gamma=0}=\sum_{k\in\mathcal{S}(h)}\lambda(\phi_k)+\frac{hL}{2}.
\end{equation}
Eq.~\eqref{gs_free} implies that if $|h|<1$ the spectrum of Eq.~\eqref{ham_gamma} is gapless in the 
thermodynamic limit $N\rightarrow\infty$ while if $|h|>1$ is gapped and the ground state is completely 
polarized. These preliminary considerations are of course very well known and already imply that all the  ground overlaps  are trivial if $|h|>1$.
Moreover, following Sec.~\ref{sec2}, it is straightforward to verify that the partition function in Eq.~\eqref{quantum_pf} becomes
\begin{equation}
\label{free_p}
 f_{\sigma}(\beta,2N)|_{\gamma=0}=e^{-\frac{\beta hL}{2}}|\det \boldsymbol{Q}_{\sigma}(\beta)|,
\end{equation}
where the matrix $\boldsymbol{Q}_{\sigma}$ is calculated from the $\gamma\rightarrow 0$ limit of $\textbf{Q}$ in Eq.~\eqref{matQ}  
by removing  rows and columns in correspondence with the positions of the down spins in the state $|\boldsymbol{\sigma}\rangle$. By 
comparing Eq.~\eqref{limit} with \eqref{free_p} and Eq.~\eqref{gs_free} one finally obtains
\begin{equation}
\label{ov_gamma0}
 \left|\langle\boldsymbol{\sigma}|\Omega\rangle|_{\gamma=0}\right|^2=\lim_{\beta\rightarrow\infty}|\det \textbf{Q}_{\sigma}(\beta)|e^{\beta\sum_{k\in\mathcal{S}(h)}\lambda(\phi_k)}.
\end{equation}
Since the matrix $\textbf{Q}$ is circulant, the proof of Eq.~\eqref{c_coeff_full}  given 
in Appendix~\ref{app2} carries over. In particular, $\det\textbf{Q}_{\sigma}$ in Eq.~\eqref{ov_gamma0} expands 
over polynomials in the eigenvalues $q(\phi_k)=e^{-\beta\lambda(\phi_k)}$, $k=1,\dots,2N$ of the matrix $\textbf{Q}$. For a 
fixed value of  the transverse field, the overlap in Eq.~\eqref{ov_gamma0} is then proportional  to the number of polynomials, 
if any,  whose value equals the Boltzmann factor of the ground state. We propose an illustrative example for the class of states 
labelled by the block $\mathcal{B}_{1,p}$; see the first example of Sec.~\ref{sec4}. From Eq.~\eqref{c_coeff_full} one has
\begin{equation}
\label{sum0}
 \det\textbf{Q}_{\sigma}(\beta)=\prod_{k=1}^{2N/p}\left(\frac{1}{p}\sum_{j=0}^{p-1}e^{-\beta\lambda(\phi_k+2\pi j/p)}\right)=\frac{1}{p^{2N/p}}\sum_{\{\boldsymbol{j}\}}e^{-\beta\sum_{k=1}^{2N/p}\lambda(\phi_k+2\pi j_k/p)},
\end{equation}
where $\boldsymbol{j}=\{j_1,\dots,j_{2N/p}\}$ with $j_i=0,\dots, p-1$. In the limit $\beta\rightarrow\infty$, the 
sum in Eq.~\eqref{sum0} is dominated by the configurations $\bar{\boldsymbol{j}}$  which minimize the $\lambda$'s---
practically $\cos(\phi_k)$---for any given value of the $\phi_k$. The extremal configuration is unique and such 
that the corresponding angles $\phi_{k}+2\pi \bar{j}_{k}/p$  cover uniformly an arc of length $2\pi/p$ centered 
around $\phi=\pi$, see Fig.~\ref{fig0} for a graphical proof. This  result implies that the ground  state overlap 
of the states labelled by $\mathcal{B}_{1,p}$ is non-zero if the arc $(\pi-\pi/p,\pi+\pi/p)$ coincides with the 
Fermi sea, namely $\arccos(h)=\pi-\pi/p$ and $\langle\mathfrak{M}\rangle/L=\frac{1-(p-1)}{p}$ as expected. Provided 
that this is the case, it is immediate to conclude that 
\begin{equation}
\label{ovp}
 \left|\langle\boldsymbol{\sigma}|\Omega\rangle|_{\gamma=0}\right|^2=\frac{1}{p^{2N/p}},
\end{equation}
and therefore $s_{\sigma}=0$, indicating renormalization toward a 
Dirichlet boundary state of a bosonic CFT~\cite{OA}. We mention that Eq.~\eqref{ovp} for $p=2$ has been also obtained in~\cite{MS}.
Finally, we have repeated the overlap calculation in Eq.~\eqref{ov_gamma0} for  the configurations $\mathcal{B}_{s,p}$ analyzed in Sec.~\ref{sec4} and in Appendix~\ref{app3}. In all the cases, the boundary entropies vanish.

\section{Conclusions}
\label{sec6}
In this paper we  studied in detail  a vast class of ground state overlaps in the XY chain when  the number 
of lattice sites is even ($L=2N$). In particular, we provided an explicit determinant 
representation,~see Eq.~\eqref{overlapXY}, adapting to imaginary times the formalism 
developed in~\cite{RA} for the Return Amplitude. From such a determinant representation 
we extracted the large $N$ limit by proving a general formula, ~see Eq.~\eqref{finalc}, for the 
principal minors of circulant matrices. The finite $O(1)$ contribution in the thermodynamic limit 
of the overlap at criticality is shown to be $\gamma$-independent for all the 
states considered. Its logarithm defines the universal renormalized Boundary Entropy~\cite{boundary}, which was proven to have 
only two possible values depending on whether the quantum state flows to the free conformal boundary state or 
to the linear superposition of fixed conformal boundary states. Linear superpositions of fixed conformal boundary states appear naturally also in the  analysis of topological defects~\cite{GW}, as a result of the Kramers-Wannier duality applied to the free  boundary state~\cite{Kon, chiral, FI}.

As already mentioned, for technical reasons, our analysis has been limited to a chain with an even 
number of lattice sites and the expressions for the ground state overlaps are  valid outside the 
circle $\gamma^2+h^2=1$. This is not a strong limitation, since the domain covers almost all the relevant critical cases~\cite{OYNR}.  For chains with an odd number of lattice sites and $h<0$ or inside the circle $\gamma^2+h^2=1$, however,  the lowest-energy state might belong to  the Ramond sector. In this case, finite-size 
corrections could develop also subleading logarithmic terms $O(\log N)$~\cite{Navot2}. It would be interesting 
to investigate this possibility and its implications in the future: the existence of logarithmic corrections to the scaling can spoil, for instance, the universality of the $O(1)$ contribution. It is also worth to further test the universality conjecture for the Boundary Entropies at the critical 
point by considering irrelevant integrability breaking interactions, such as next-to-next neighbour couplings. Overlaps 
can be calculated numerically through Matrix Product approximations of the ground state.

Finally, we have also discussed the case $\gamma=0$, where Formation Probabilities are  directly related to 
the multiplicity of the ground state Boltzmann weight in the large $\beta$ expansion of a suitable 
determinant, see Eq.~\eqref{ov_gamma0}. In this case, our study  extends considerably the analytic results for the overlaps presented in~\cite{MS}.

\section*{Acknowledgements}
MAR thanks CNPq and FAPERJ (grant number 210.354/2018) for partial support. FA and JV are partially supported by the Brazilian Ministries MEC and MCTC, the CNPq (grant number 306209/2019-5) and the Italian Ministry MIUR under the grant PRIN 2017  ``Low-dimensional quantum systems: theory, experiments and simulations".

\begin{appendix}
\section{Euler-MacLaurin (EM) summation formulas}
\label{app}
\noindent
\textit{EM Summation Formula.---}If $f(x)$ is a differentiable function in the interval $0\leq x\leq 1$, then
\begin{equation}
\label{EM1}
 \sum_{k=1}^N f\left(\frac{k-1/2}{N}\right)=N\int_{0}^1 dx f(x) 
 -\frac{f'(1)-f'(0)}{24}\left(\frac{1}{N}\right)+O(1/N^2),
\end{equation}
see Eq. (1) in~\cite{Navot1}  for $a=1/2$.
In~\cite{Navot2}, an extension of the EM summation formula  was proven, of which we made extensive application in this paper.\\

\textit{Extended EM Summation Formula.---} Take an integrable function $f(x)$, in the interval $0\leq x\leq 1$, such that $f(x)\stackrel{x\rightarrow 0}{\rightarrow}\log x^{\alpha}$, $\alpha\in\mathbb R$. Then the following summation formula holds
\begin{equation}
\label{EM2}
 \sum_{k=1}^N f\left(\frac{k-1/2}{N}\right)=N\int_{0}^1 dx f(x)+\frac{\alpha \log 2}{2}+O(1/N),
\end{equation}
 see Eq. (7) in~\cite{Navot2} for $g(x)=\alpha$ and $a=1/2$.
Notice that differently from Eq.~\eqref{EM1}, the $O(1)$ term is now non-zero. If more than one logarithmic singularity is present on the integration domain, it is always possible to divide it in subsets such that any subset will contain only one singularity. It is clear than that the contributions of different singularities add up.

For the sake of completeness, we provide a quick but non rigorous proof of Eq.~\eqref{EM2}. Let us consider $f(x)$ as above and rewrite $f(x)=g(x)+\alpha\log(x)$, where  $g(x)\equiv f(x)-\alpha\log(x)$ satisfies the hypothesis of the EM Summation Formula, Eq.~\eqref{EM1}. Proving the extended EM Summation Formula boils down to estimate the large $N$ limit of the sum
\begin{equation}
\sum_{k=1}^{N}\log\left(\frac{k-1/2}{N}\right),
\end{equation}
which can be done by expanding its exponential, i.e. $\frac{(2N-1)!!}{(2N)^N}$, for $N\gg 1$. By applying the Stirling formula one obtain $\frac{(2N-1)!!}{(2N)^N}=\sqrt{2}e^{-N+O(1/N)}$ from which Eq.~\eqref{EM2} easily follows~\cite{Ref2}. 

\section{Proof of Eq.~\eqref{c_coeff_full}}
\label{app2}
In order to determine the overlap in Eq.~\eqref{overlapXY} and especially the determinant of the $(sM)\times (sM)$ matrix $\textbf{W}_{\sigma}$, we 
proceed as follows. Let us introduce a $2N\times 2N$ diagonal matrix $\textbf{I}^{(sp)}$, with 
elements $[\textbf{I}^{(sp)}]_{lm}=\delta_{l,m}\delta_{l,q}$, being $q$ the position of an up spin 
in the configuration labelled by $\mathcal{B}_{s,p}$. The matrix $\textbf{W}'_{\sigma}=\textbf{W}\textbf{I}^{(sp)}$ 
will have rank $sM$ and columns of zeros in correspondence with the positions of the down spins. Consider now the characteristic polynomial 
\begin{equation}
\label{car_pol}
 \mathfrak{P}_M\equiv\det(\lambda\textbf{I}-\textbf{W}_{\sigma}')=\sum_{n=0}^{2N} c_{n}\lambda^n.
\end{equation}
It is known, see for example~\cite{Matan}, that its coefficients $c_{n}$ can be expressed in terms of the 
principal minors of order $2N-n$ of the matrix $\textbf{W}_{\sigma}'$. We recall for convenience that a 
principal minor of order $2N-n$ of a $2N\times 2N$ matrix is the determinant of the $(2N-n)\times (2N-n)$ sub-matrix 
obtained by removing the \textit{same} set of  $n$ rows and columns from the original matrix. It then follows from the 
previous considerations and the definition of the matrix $\textbf{W}_{\sigma}$ in Eq.~\eqref{overlapXY} that the first 
non-vanishing coefficient of the characteristic polynomial in Eq.~\eqref{car_pol} is $c_{2N-sM}=c_{M(p-s)}$ and moreover
\begin{equation}
\label{finalc}
 c_{M(p-s)}=\det \textbf{W}_{\sigma}.
\end{equation}
We now discuss how the coefficient $c_{M(p-s)}$ of the characteristic polynomial in Eq.~\eqref{car_pol} can be calculated in closed form.

The matrix $\textbf{I}^{(sp)}$, entering  the definition of $\textbf{W}_{\sigma}'$, reads (see for instance~\cite{LC}) 
\begin{equation}
\label{ex_I}
 [\textbf{I}^{(sp)}]_{lm}=\frac{\delta_{lm}}{p}\sum_{j=0}^{p-1}\sum_{r=0}^{s-1}e^{\frac{2\pi ij(l+r)}{p}}.
\end{equation}
Notice also that the matrix $\textbf{W}$ is circulant (cf Eq.~\eqref{wlim}) and can be diagonalized by the unitary matrix $[\textbf{U}]_{lk}=\frac{1}{\sqrt{2N}}e^{i l\phi_k}$, in particular $\textbf{D}_{W}\equiv\textbf{U}\textbf{W}\textbf{U}^{\dagger}=\text{diag}(w(\phi_1),\dots, w(\phi_{2N}))$. Because of the form of the states chosen in Sec.~\ref{sec4}, see Eq.~\eqref{M-def}, $2N=Mp$.
To express the coefficient $c_{M(p-s)}$   of the characteristic polynomial of the matrix $\textbf{W}_{\sigma}'$ in terms of the eigenvalues of $\textbf{W}$, it is convenient to rewrite
\begin{equation}
\label{car_pol_2}
 \mathfrak{P}_M(\lambda; \mathcal{W})=\det(\lambda \textbf{I}-\overbrace{\textbf{D}_{W}\textbf{U}\textbf{I}^{(sp)}\textbf{U}^{\dagger}}^{\textbf{A}_M}),
\end{equation}
where $\mathcal{W}=\{w(\phi_1),\dots, w(\phi_{Mp})\}$ and we made evident all the variable dependence. In the rest of the Appendix, we will further use the shorthand notation $w_l$  for $w(\phi_l)$.
The matrix $\textbf{A}_M$ can be calculated explicitly from its definition in Eq.~\eqref{car_pol_2} and one finds
\begin{equation}
\label{matA1}
 [\textbf{A}_M]_{lm}=\frac{w_l}{p}~\delta_{l,m}^{\text{mod}M}\sum_{j=p-s+1}^{p}e^{\frac{2\pi i(j-1/2)(l-m)}{Mp}},
\end{equation}
with $l,m=1,\dots, Mp$. However, since $l-m=Mk~(k=0,\dots,p-1)$  as a consequence of the Kronecker symbol, Eq.~\eqref{matA1} simplifies to
\begin{equation}
\label{matA2}
 [\textbf{A}_M]_{lm}=w_l~\delta_{l,m}^{\text{mod}M}B_{k}\quad\text{where}\quad B_{k}\equiv\frac{1}{p}\sum_{j=p-s+1}^{p}e^{\frac{2\pi i(j-1/2)k}{p}}.
\end{equation}
Eq.~\eqref{matA2} implies  that the coefficients $B_{k}$ are $M$-independent, moreover it is easy to verify that $\textbf{A}_1$ coincides with $\textbf{A}$ in Eq.~\eqref{matA}, replacing $w_l\leftrightarrow x_l$. By denoting with $[\bullet]_{\lambda}=\lambda\cdot 1-\bullet$, the characteristic polynomial of $\textbf{A}_1$ is 
\begin{equation}
\label{car_pol_ap}
 \mathfrak{P}_1(\lambda; \{w_1,\dots,w_p\})=\det\begin{bmatrix}[w_1B_0]_\lambda& -w_1B_1 & -w_1B_2 &\dots & -w_1B_{p-1} \\
 -w_2 B_1^*&[w_2B_0]_{\lambda} & -w_2B_1 &\dots & -w_2B_{p-2} \\
 \vdots & \vdots & \vdots & \vdots & \vdots\\
-w_p B_{p-1}^* & -w_p B_{p-2}^* & -w_pB_{p-3}^* &\dots & [w_pB_0]_{\lambda} \\
                         \end{bmatrix}.
\end{equation}
To demonstrate Eq.~\eqref{c_coeff_full} one proceeds by induction on $M$. First, we will prove that the characteristic polynomial of the matrix $\textbf{A}_M$ is factorized, that is
\begin{equation}
\label{ind}
 \mathfrak{P}_{M}(\lambda; \mathcal{W})=\prod_{k=1}^{M}\mathfrak{P}_{1}(\lambda; \mathcal{W}_k),
\end{equation}
where each of the $M$ sets $\mathcal{W}_k$  contain the  $p$ variables $w_{k+jM}$ for $j=0,\dots,p-1$.

A moment of thought shows that $\textbf{A}_M$ is actually $\textbf{A}_1$ with the property that elements on different diagonals have been separated by $M-1$ diagonals of zeros; therefore its characteristic polynomial $\mathfrak{P}_M$ is 
\begin{equation}
\label{bigmat}
 \det\left\lceil
 \begin{matrix} 
[w_1B_0]_{\lambda} & 0 &\dots & 0 & -w_1B_1 & 0 & \dots & 0 & -w_1B_2 \\
0 & [w_2B_0]_{\lambda} & 0  &\dots & 0 &-w_2 B_1 & 0 & \dots & 0\\ 
\vdots & \vdots & \vdots  &\vdots & \vdots &\vdots  & \vdots & \vdots & \vdots\\
0 & \dots & 0 & [w_{M}B_0]_{\lambda} & 0 & \dots & 0& -w_M B_1 & 0 \\
-w_{M+1} B_1^* & 0 & \dots & 0 & [w_{M+1} B_0]_{\lambda} & 0 & \dots & 0  & -w_{M+1} B_1\\
\vdots & \vdots & \vdots  &\vdots & \vdots &\vdots  & \vdots & \vdots & \vdots
\end{matrix}\right..
\end{equation}
Ignoring signs that can be easily traced back, it is possible to move columns and rows of the matrix in Eq.~\eqref{bigmat} to calculate its determinant. We then accommodate to the left, after $\frac{(M-1)p(p-1)}{2}$ exchanges, all the columns that contain the variables $w_1, w_{1+M}, w_{1+2M},\dots, w_{1+(p-1) M}$. Dropping a factor  $(-1)^{\frac{(M-1)p(p-1)}{2}}$, one  ends up with the following  expression for $\mathfrak{P}_M$
\begin{equation}
\label{bigmat2}
\det\left\lceil\begin{matrix}
[w_1 B_0]_{\lambda} & -w_1B_1 & \dots &-w_1 B_{p-1} & 0 & 0 &  0 & \dots & 0 \\
0 & 0 &\dots & 0 & [w_2 B_0]_{\lambda} & 0 & \dots & 0 &-w_2 B_1\\
0 & 0 &\dots & 0 & 0 & [w_3B_0]_{\lambda} & 0 & \dots &0  \\
\vdots & \vdots &\vdots & \vdots & \vdots & \vdots & \vdots & \vdots &\vdots\\
-w_{M+1} B_1^* & [w_{M+1}B_0]_{\lambda} &\dots & -w_{M+1} B_{p-2} & 0 & 0 &  0 & \dots & 0\\
0 & 0 &\dots & 0 & -w_{M+2} B_1^{*} & 0 & \dots & 0 &[w_{M+2} B_0]_{\lambda}\\
\vdots & \vdots & \vdots  &\vdots & \vdots &\vdots  & \vdots & \vdots & \vdots
\end{matrix}\right..
\end{equation}
  By pushing up, with again $\frac{(M-1)p(p-1)}{2}$ exchanges, all the rows labelled by $1+jM$ with $j=1,\dots, p-1$ of Eq.~\eqref{bigmat2}, we finally arrive at
\begin{equation}
\label{mat3}
 \mathfrak{P}_{M}=\det\begin{bmatrix} [\textbf{A}_1(\mathcal{W}_1)]_{\lambda} & \textbf{0} \\
 \textbf{0} & [\textbf{A}_{M-1}(\overline{\mathcal{W}}_1)]_{\lambda} \\
                             \end{bmatrix}.
\end{equation}
The notation in Eq.~\eqref{mat3} indicates that the matrix $\textbf{A}_{1}$ contains all the variables in the set $\mathcal{W}_1=\{w_1,\dots,w_{M(p-1)+1}\}$ organized in the same order as in Eq.~\eqref{car_pol_ap}.The matrix $\textbf{A}_{M-1}$ is instead a function of the remaining variables; namely $\overline{\mathcal{W}}_1=\mathcal{W}\backslash\mathcal{W}_1$. We have then proven that
\begin{equation}
\label{almost_f}
 \mathfrak{P}_{M}(\lambda;\mathcal{W})=\mathfrak{P}_1(\lambda; \mathcal{W}_1)
  \mathfrak{P}_{M-1}(\lambda; \overline{\mathcal{W}}_1).
\end{equation}
By applying the inductive hypothesis to $\mathfrak{P}_{M-1}$ in Eq.~\eqref{almost_f}, the first part of our proof, i.e. Eq.~\eqref{ind}, now follows. It is left to show that the coefficient of the lowest power of $\lambda$ in  $\mathfrak{P}_M(\lambda; \mathcal{W})$ is also factorized. The matrix $\textbf{A}_1$ has rank $s$ and  $p-s$ among its eigenvalues are zero. Let $\mathcal{P}_{j}(x_1,\dots,x_p)$ with $j=0,\dots,s$ denote the coefficients of $\lambda^{p-s+j}$ in the characteristic polynomial $\mathfrak{P}_1(\lambda; \{x_1,\dots,x_p\})$. The latter are determined recursively by the Faddeev-Le Verrier algorithm~\cite{FF};  for example: $\mathcal{P}_{s}=1$ and $\mathcal{P}_{s-1}=-\text{Tr}[\textbf{A}_1]$. From Eq.~\eqref{ind} we thus conclude that 
\begin{equation}
\label{pol}
 \mathfrak{P}_M(\lambda; \mathcal{W})=\lambda^{M(p-s)}\prod_{k=1}^{M}\left[\mathcal{P}_s(w_{k},\dots,w_{k+(p-1)M})\lambda^s+\dots+\mathcal{P}_0(w_{k},\dots,w_{k+(p-1)M})\right],
\end{equation}
where we have made explicit the variable dependence of the  polynomials $\mathcal{P}_{j}$. The lowest power of $\lambda$ in Eq.~\eqref{pol} is $M(p-s)$ and its coefficient is
\begin{equation}
\label{app_fin}
 c_{M(p-s)}=\prod_{k=1}^{M}\mathcal{P}_0(w_{k},\dots,w_{k+(p-1)M}),
\end{equation}
eventually proving Eq.~\eqref{c_coeff_full}. Notice that the result in Eq.~\eqref{app_fin} holds for any circulant matrix $\textbf{W}$.

\section{Additional Examples}
\label{app3}
In the final Appendix, we gather additional examples of calculations of the BE for states labelled by the blocks $\mathcal{B}_{s,p}$ at $\gamma\not=0$. The results are summarized in Tab.~\ref{tab1}, where we provide the polynomial $\mathcal{P}_0(x_1,\dots,x_p)$, see Eq.~\eqref{overlap_fin}, and the values of the BE along the critical lines $h=\pm 1$. The latter, as explained in many occasions in the main text, are obtained by analyzing the zeros and singularities in the domain $\phi\in[0,2\pi/p]$ of the function
\begin{equation}
 g_{s,p}(\phi)\equiv|\mathcal{P}_0\bigl(w(\phi),w(\phi+2\pi/p),\dots,w(\phi+2\pi(p-1)/p)\bigr)|,
\end{equation}
and applying Eq.~\eqref{EM2}.
\begin{table}[t]
\begin{tabular}{|c|c|c|c|}
\hline
$\mathcal{B}_{s,p}$ & $\mathcal{P}_0(x_1,\dots,x_p)$ & $s_{\sigma}(1)$ & $s_{\sigma}(-1)$\\
\hline
$\mathcal{B}_{2,3}$ & \tiny{$\frac{1}{3}(x_1x_2+x_1x_3+x_2x_3)$} & $\frac{1}{2}\log 2$ & 0\\
$\mathcal{B}_{2,5}$ & \tiny{$\frac{1}{50} \left(x_1 \left(\eta  x_3+\eta  x_4+\xi  x_2+\xi  x_5\right)+x_2 \left(\eta 
   \left(x_4+x_5\right)+\xi  x_3\right)+\eta  x_3 x_5+\xi  x_4
   \left(x_3+x_5\right)\right)$}  & $\frac{1}{2}\log 2$ & 0\\
$\mathcal{B}_{3,5}$ & \tiny{$\frac{1}{50} \left(x_2 \left(x_4 \left(-\eta  x_5-\xi  x_3\right)-\eta  x_3
   x_5\right)+x_1 \left(x_2 \left(-\eta  x_4-\xi  x_3-\xi  x_5\right)-\eta  x_3
   \left(x_4+x_5\right)-\xi  x_4 x_5\right)-\xi  x_3 x_4 x_5\right)$} & 0 & $\frac{1}{2}\log 2$\\
$\mathcal{B}_{4,5}$ & \tiny{$\frac{1}{5} \left(x_2 x_3 x_4 x_5+x_1 \left(x_3 x_4 x_5+x_2 \left(x_3
   x_4+\left(x_3+x_4\right) x_5\right)\right)\right)$} & $\frac{1}{2}\log 2$ & 0\\
\hline
 \end{tabular}
 \caption{Results for the BEs up to $p=5$, $s_{\sigma}=0$ corresponds to the free boundary state, while $s_{\sigma}=\frac{1}{2}\log 2$ indicates renormalization toward the linear superposition of fixed boundary states. In the Table, we have defined
 $\eta=5+\sqrt{5}$, $\xi=5-\sqrt{5}$. \texttt{Mathematica} gives explicit expressions for the polynomials also for larger values of $p$ but they become increasingly cumbersome. The values of the BEs calculated from Eq.~\eqref{EM2} are consistent with the general pattern enunciated at the end of Sec.~\ref{sec4}. }
 \label{tab1}
 \end{table}
\end{appendix}

\end{document}